\documentclass[showpacs,pre,aps, twocolumn]{revtex4-1}
\usepackage{graphicx}
\usepackage{xcolor}
\usepackage{dcolumn}
\usepackage{bm}
\def\gapx{\lower 2pt \hbox{$\buildrel>\over{\scriptstyle{\sim}}$\ }}
\def\lapx{\lower 2pt \hbox{$\buildrel<\over{\scriptstyle{\sim}}$\ }}

\RequirePackage[colorlinks,linkcolor=black,citecolor=blue,hyperindex]{hyperref}

\begin{document}
\title{Monodisperse cluster crystals: classical and quantum dynamics}
\author{Rogelio D\'iaz-M\'endez,$^{1}$ Fabio Mezzacapo,$^1$ Fabio Cinti,$^2$ Wolfgang Lechner,$^{3}$}
\author{Guido  Pupillo$^1$}
\affiliation{$^1$IPCMS (UMR 7504) and ISIS (UMR 7006), Universit\'e de Strasbourg and CNRS, Strasbourg, France\\
$^2$National Institute for Theoretical Physics (NITheP), Stellenbosch 7600, South Africa\\
$^3$IQOQI and Institute for Theoretical Physics, University of Innsbruck, Austria}

\date{\today}

\begin{abstract}
We study the phases and dynamics of a gas of monodisperse particles interacting via soft-core potentials in two spatial dimensions, which is of interest for soft-matter colloidal systems and quantum atomic gases. Using exact theoretical methods, we demonstrate that the equilibrium low-temperature classical phase simultaneously breaks continuous translational symmetry and dynamic space-time homogeneity, whose absence is usually associated with out-of-equilibrium glassy phenomena. This results in an exotic self-assembled cluster crystal with coexisting liquid-like long-time dynamical properties, which corresponds to a classical analog of supersolid behavior. We demonstrate that the effects of quantum fluctuations and bosonic statistics on cluster-glassy crystals are separate and competing: zero-point motion tends to destabilize crystalline order, which can be restored by bosonic statistics.
\end{abstract}


\maketitle
\section*{INTRODUCTION}
The discovery of novel phases of matter as a result of broken symmetries is of main interest in condensed matter. In quantum physics, a key example is the supersolid phase for bosonic particles \cite{Kuklov2011, Boninsegni2012, Chan2013}, where the rare simultaneous breaking of two symmetries (i.e., continuous translational and global gauge symmetry) leads to the coexistence of both crystalline and superfluid properties \cite{Gross1957,Saccani2012, Henkel2010,Cinti2010,Henkel2012,Cinti2014, Danshita2009, Stringari2013}. Such state of matter may be realized for a monodisperse ensemble of particles with  cluster forming interactions, at sufficiently low temperatures~\cite{Cinti2014,Cinti2010}. 

A phase that does not fall into the scheme of broken symmetries is the glass phase, both classical and quantum, which is a non-equilibrium and disordered - yet stable - phase \cite{Merolle2005,ScottShell2005}. In analogy to the broken-symmetry picture, however, the glass transition is often associated with breaking of space-time homogeneity, or {\it dynamic heterogeneity}. The latter is a result of frustration effects, known as self-caging \cite{ZAMPONI}, and corresponds to a relaxation that is fast on a local scale, and exponentially slow on large-ones  \cite{BINDERBOOK,WOLYNESBOOK,DEBENEDETTI2001,REVWEEKS,SCORTINO,TANAKACRITICAL,GOETZE}. The search for novel mechanisms for caging and glass effects both in the classical \cite{Sciortino2013} and quantum  \cite{Zamponi2011,Biroli2008, Nussinov2008,Carleo2009, QUANTUMJAMMING,Nussinov2013} regimes is of central interest in condensed matter, as well as atomic and molecular physics \cite{LZ2013,igor}. Interestingly, several works have pointed out that mixtures of cluster forming particles  {\it out-of-equilibrium} may realise glassy phases in the classical regime~\cite{Sciortino2013,Zamponi2011}. This opens the way to understand  the relation between the equilibrium and
non-equilibrium properties of classical cluster-forming particles compared to their quantum counterparts~\cite{Sciortino2013},
and thus to investigate the classical to quantum transition in these systems.
Indeed, a key question is whether glassiness and supersolidity may be related at a fundamental level.

In this work we  analyze the classical {\it equilibrium} phases, and the effects of quantum fluctuations and bosonic statistics on such phases, in a two dimensional  model system of ultra-soft particles that, in the quantum regime, has been shown to display the existence of a bosonic cluster supersolid~\cite{Cinti2010,Henkel2010,Henkel2012,Cinti2014}. In such a state of matter superfluidity emerges in  the presence of a self-assembled cluster-crystalline structure. Here we focus on the classical counterpart of this quantum phase, and investigate in detail both its static and dynamical properties using exact theoretical techniques that are valid in the classical and semiclassical regimes. 

Specifically, we demonstrate theoretically that (i) the low-temperature equilibrium classical phase is an exotic ordered cluster crystal with a dynamical separation between intra-cluster particle motion and inter-cluster hopping, mimicking  dynamic heterogeneity  {\it at equilibrium}. This is microscopically due to caging effects at the level of individual clusters, which result in the coexistence of crystalline and liquid-like properties, such as linear particle diffusion as a function of time $t$. This clarifies in what sense this thermodynamic phase is the classical version of a cluster supersolid. (ii) By means of a combination of numerically exact semiclassical and fully quantum techniques we elucidate the different effects of quantum fluctuations and quantum statistics on the  phenomena described here. Surprisingly, we find that these are competing: zero-point motion tends to destabilize the cluster crystal in favor of liquid-like phases,
while bosonic quantum statistics can have the opposite effect of enhancing crystalline behavior. 
The latter effect is at  odds with results for systems such as He$^4$ or dipolar crystals \cite{QUANTUMJAMMING}, where bosonic statistics always favors
a liquid behavior, while the former  should be compared to recent findings for  polydispersed hard spheres \cite{MARKLAND2011,Zamponi2011,Nussinov2013}, where  quantum fluctuations  may help
crystallization. Some of these effects may be relevant for systems as diverse as colloidal particles
\cite{Li2004, Narros2010, Sciortino2013} as well as cold gases of Rydberg atoms~\cite{exp}, where light-dressing techniques can be used to tune effective inter particle interactions~\cite{Santos2001,Biedermann2014}.\\ 

The remainder of this paper is organized as follows: in the next Sec.~\ref{Sec:Model} we present the Hamiltonian of interest for an ensemble of particles interacting via ultrasoft, cluster-forming potentials and summarise known results in the quantum regime. In Sec.~\ref{Sec:PD} we present the computed classical phase diagram for our model, by introducing both static and dynamical order parameters which allow us to characterize the classical counterpart of the quantum cluster supersolid investigated in Ref. \cite{Cinti2014}.  In Sec.~\ref{Sec:Quantum} we introduce exact results for the semiclassical real-time dynamics of our model, by excluding particle exchanges, as well as exact results for static properties that include the effects of bosonic quantum statistics. In this way, namely by understanding the separate effect of zero-point motion and bosonic quantum statistics on the classical phase, we start elucidating the transition between the classical and quantum scenario. Finally,  we outline the conclusions as well as possible extensions of the present work.

\section{Hamiltonian: Ultrasoft particles}\label{Sec:Model}
We consider a two-dimensional ensemble of $N$ bosonic particles with mass $m$, density $\rho$ and  Hamiltonian 
\begin{equation}\label{eq:ham}
\hat{H}=-\frac{\hbar^2}{2m}\sum_{i=1}^{N}\nabla^2_i+\sum_{i<j}^N\frac{V_0}{r_{ij}^\gamma+R_{\rm c}^\gamma}.
\end{equation}
The interaction in Eq.~(\ref{eq:ham}) approaches a constant value $V_0/R_{\rm c}^\gamma$ as the inter-particle
distance $r$ decreases below the soft-core distance $R_{\rm c}$, and drops to zero for $r>R_{\rm c}$. The case
$\gamma\rightarrow\infty$ yields the soft-disc model 
\cite{Pomeau1994}. Here  we focus on $\gamma=6$, corresponding to soft-core van der Waals interactions of relevance for
ultracold atoms 
\cite{Henkel2010,mhs11}.

\indent Particles with soft-core interactions have been studied previously 
\cite{LIKOS2001,mgk06, likos2007, csk11,k2013} in the classical high-temperature regime ($\hbar = 0$, $T\neq 0$), and in the purely quantum
zero-temperature regime  ($\hbar\neq 0$, $T = 0$) 
\cite{Henkel2010,Cinti2010,Cinti2014}. In the former case, it has been
shown that pair potentials with a negative Fourier component 
\cite{mgk06} favor the formation of particle clusters, which
in turn can crystallize to form so-called classical cluster-crystals. While cluster formation has been intensively investigated in the context of e.g., colloids  \cite{COL}, it has been demonstrated that bosonic quantum statistics can turn
a cluster-solid into a supersolid phase via a quantum phase transition at a critical value $\alpha_{cs-ss} \simeq 40$ where
 $\rho mV_0/(\hbar^2 R_c^2)\equiv \alpha$ 
 \cite{Cinti2014}. The supersolid further melts into a  superfluid phase via a first
order quantum phase transition at $\alpha_{ss-sf} \simeq 30$.\\

Here, we bridge the gap between the classical and quantum
regimes by first analyzing the static and dynamic properties at the classical level 
and then studying the separate effects of quantum fluctuations and statistics. We performed our theoretical investigation by using different exact numerical approaches that are appropriate for the various regimes of our interest. Specifically, Langevin molecular dynamics \cite{chaikin95} has been employed to obtain the classical phase diagram of model Eq.~(\ref{eq:ham}). Quantum effects on the classical phases have been studied by means of Path Integral Molecular Dynamics in the semiclassical regime \cite{MARKLAND2011} and full Path Integral Quantum Monte Carlo simulations \cite{Boninsegni2006}. The former allows for investigating the real-time dynamics of quantum particles in the absence of quantum statistics, while the latter provides exact results for static properties of bosonic quantum systems.

\section{Classical phase diagram}\label{Sec:PD}
\subsection{Static properties}
\begin{figure}[t]
\centerline{\includegraphics[width = 1.0\columnwidth]{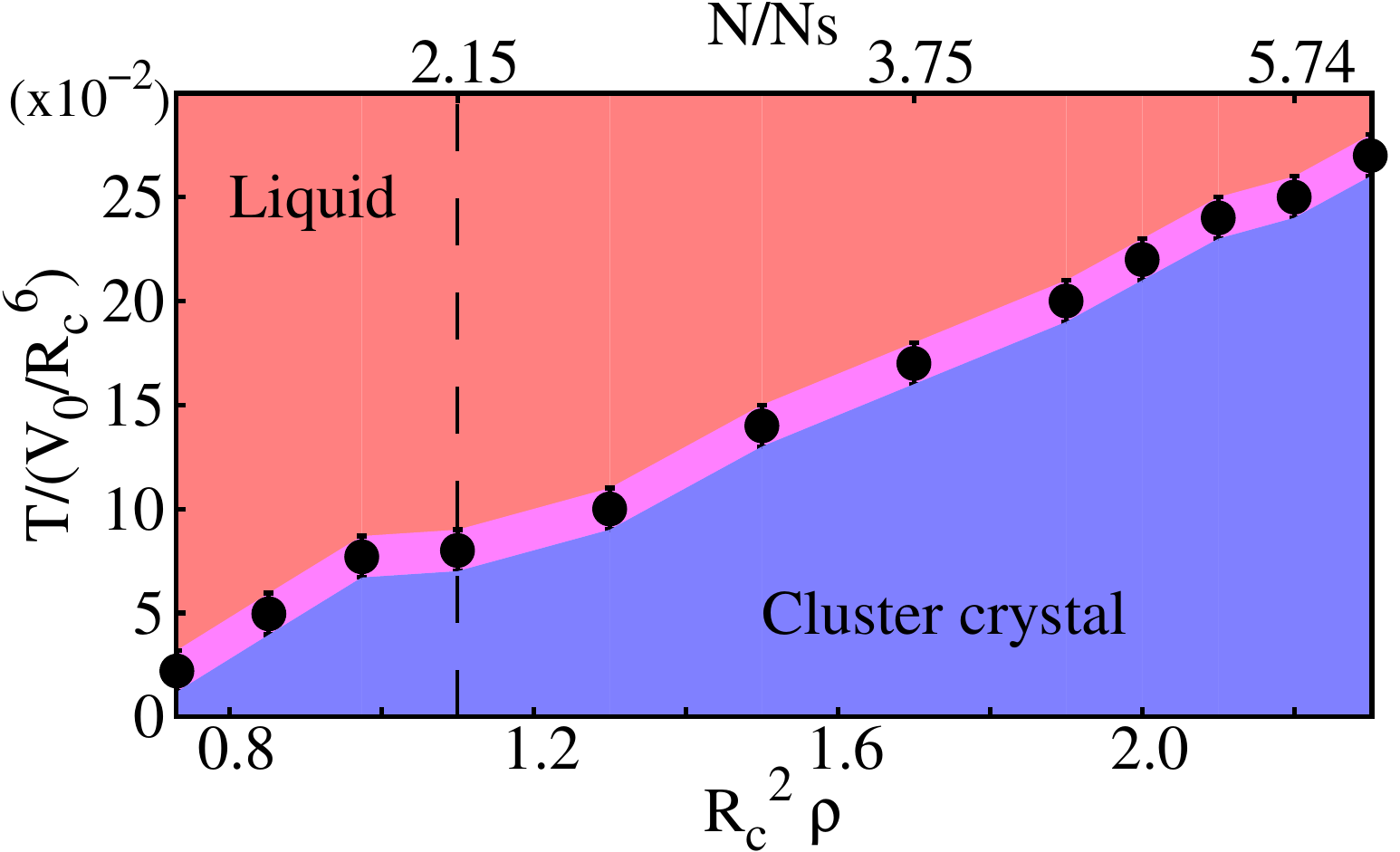}}
\caption{(color online) Phase diagram as a function of scaled temperature $T/(V_0/R_c^6)$  and  density $R_c^2\rho$  for
classical monodisperse particles with soft-core interactions [see Eq.~(1)] at equilibrium. $N/N_s$ is the average number of particles per cluster in the ground-state [upper $x$-axis]. The dashed vertical line marks $R_c^2\rho \simeq 1.1$ [see Fig. \ref{fig:fig2}].}. 
\label{fig:PD}
\end{figure}

\begin{figure}[b]
\centerline{\includegraphics[width = 1.0\columnwidth]{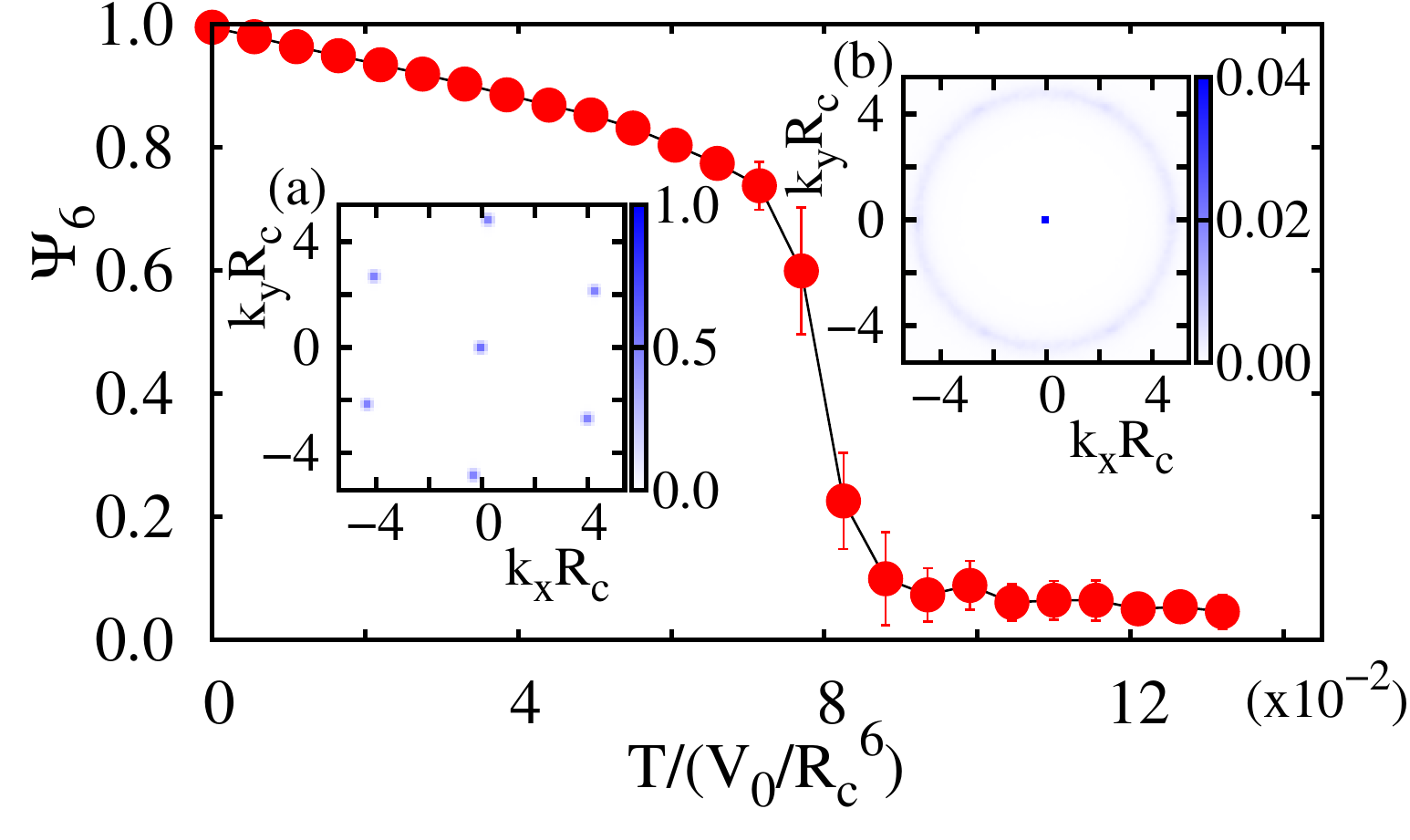}}
\caption{(color online) Cluster short-range orientational order $\Psi_6$ as a
function of $T$ at equilibrium for $R_c^2\rho \simeq 1.1$. Insets: Static structure factor $S(\mathbf k)$ at the same rescaled density for $T/(V_0/R_c^6)=0.02$ (a) and $T/(V_0/R_c^6)=0.09$ (b). In the former case the system is a cluster crystal, in the latter, as indicated by the extremely small value of  $\Psi_6$, and the essentially featureless $S(\mathbf k)$, a liquid.} 
\label{fig:fig2}
\end{figure}

The classical ($\hbar = 0$, $T\neq 0$) phases of model Eq.~(\ref{eq:ham}), as obtained from Langevin molecular dynamics \cite{chaikin95}, have been characterized  by analyzing  both static and dynamical physical observables for systems comprising up to $N=3120$ particles in a wide range of temperatures and densities. Here, thermodynamic equilibrium has been achieved by means of a standard annealing procedure.
The resulting {\it equilibrium} phase diagram is shown in Fig.~\ref{fig:PD} as a function of  temperature $T / (V_0/R_c^6)$ and the rescaled
density $R_c^2\rho \gtrsim 1$. At  low  $T$  we find an ordered crystalline phase consisting of clusters arranged in a triangular configuration, each cluster comprising an average number of particles larger than one (see e.g., Figs. \ref{fig:fig4} and \ref{fig:fig7}) and increasing with
$R_c^2\rho$. Long-range positional order has been first measured  by estimating the static structure factor
\begin{equation}
S(\mathbf k)=\frac{1}{N}\langle|\sum_{j}^N
 e^{i\mathbf{k}\cdot\mathbf{r}_j}|^2\rangle.
 \label{eq:sk}
\end{equation}
Here, $\mathbf{k}$ is a wave vector, $\mathbf{r}_j$ the position of the $j$-th particle, and $\langle \cdots \rangle$ denotes averaging over many configurations. This observable displays well defined peaks in the low $T$ ordered phase [Fig.~\ref{fig:fig2}, inset (a)], while, as expected, for high $T$ we find a normal liquid phase with no peaks in $S(\mathbf k)$ for $\mathbf k\neq0$, independently of  $R_c^2\rho$ [Fig.~\ref{fig:fig2}, inset (b)].\\

\begin{figure}[b]
\centerline{\includegraphics[width=1.0\columnwidth]{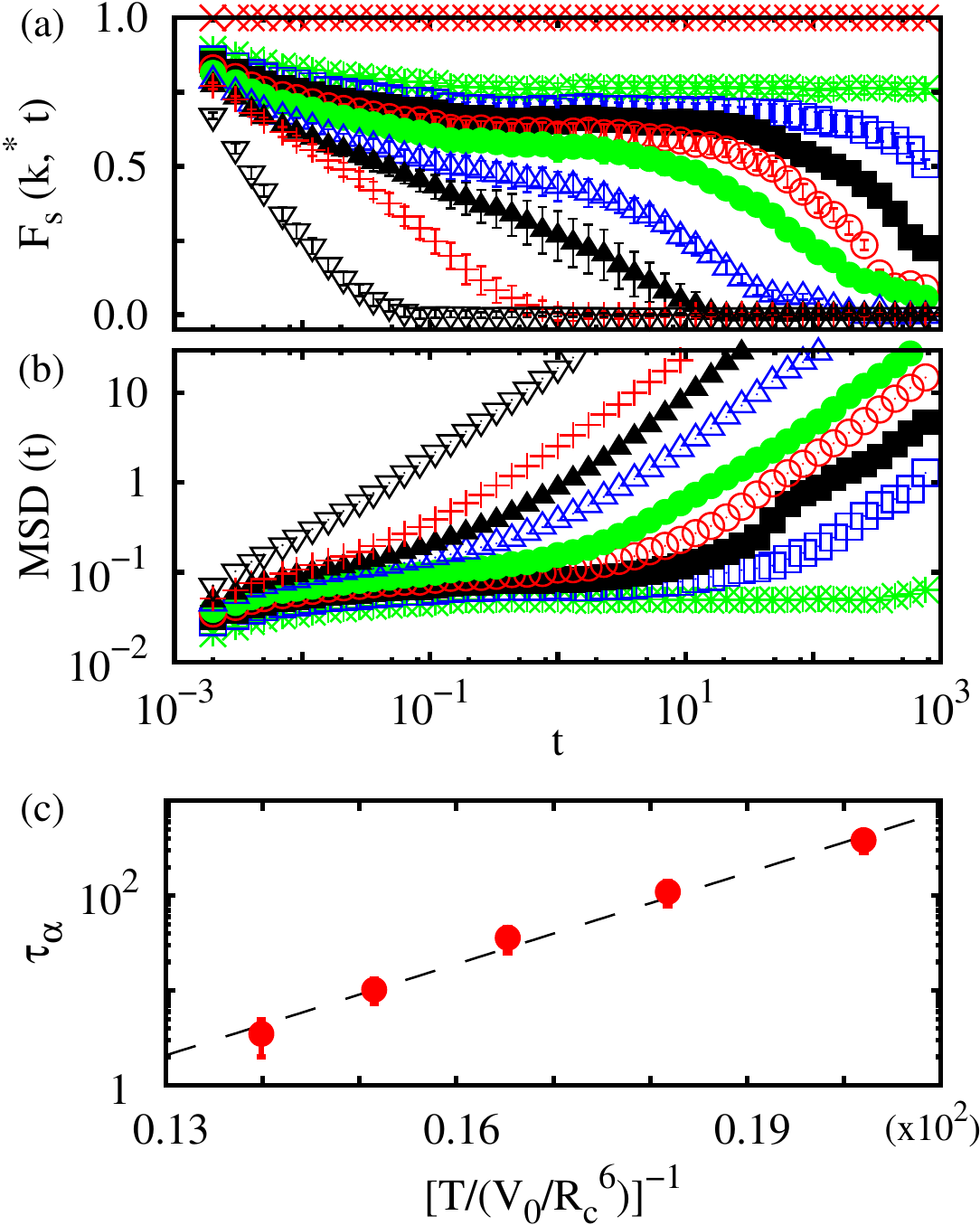}}
\caption{(color online) (a) self-intermediate scatter function  $F_s(k^*,t)$ $vs.$  time $t$ for 
$T / (V_0/R_c^6) \times 10^{2}=$ 0, 3.3, 4.4, 4.95, 5.5, 6.05, 7.15,7.70, 8.25, and 13.2  (top to bottom). (b): mean square displacement (in units of $R_c^2$) $MSD(t)$ $vs.$ $t$; symbols are the same as above.
(c): relaxation time $\tau_\alpha$ defined via $F_s(k^*,\tau_\alpha)=e^{-1}$. The dashed line is an exponential
fitting function (see text). 
Data refer scaled density $R_c^2\rho=1.1$.}
\label{fig:fig3}
\end{figure}

We then characterize quantitatively the finite-$T$ melting transition between the cluster crystal and the normal liquid phase by monitoring the hexatic (short-range) bond-order parameter of the clusters, which, in analogy to regular non-cluster forming crystals (see \cite{Note0} and Appendix for details),  we define as
\begin{equation} 
\Psi_6=\frac{1}{N_c N_j}\langle|\sum_j^{N_c}\sum_l^{N_j}e^{i6\theta_{jl}}|\rangle.
\label{eq:hex}
\end{equation}
Here, $N_c$ is the  total number of clusters, $N_j$ is the number of clusters neighboring the $j$-th one, and  $\theta_{jl}$ is  the angle between a reference axis and the segment joining the clusters $j$ and $l$ (Fig.~\ref{fig:fig7} in the Appendix).
The parameter $\Psi_6(T)$ decreases from $\Psi_6=1$ at $T=0$ to $\Psi_6\simeq 0$ in the liquid phase, and displays a sudden jump at the transition point  (see the main panel of Fig.~\ref{fig:fig2} for an example). This observed jump  is system size independent for $N\gtrsim200$ and consistent with a first order transition. 
We find that for $R_c^2\rho \gtrsim 1$ the classical melting
temperature $T_{\rm M}$ grows essentially linearly with $R_c^2\rho$, 
with a scaling of the critical interaction strength $\alpha_{cc-l}\equiv \rho V_0/(T_{\rm M}R_c^4)\simeq 0.16$.\\ 

The precise measurement of $\Psi_6$ allows us to obtain the complete phase diagram of Fig.~\ref{fig:PD} by monitoring {\it static} properties within each phase. In the next section we show that dynamical observables are necessary in order to fully characterise the various phases of the model at equilibrium. 
\begin{figure}[t]
\includegraphics[width= 1\columnwidth, angle=0]{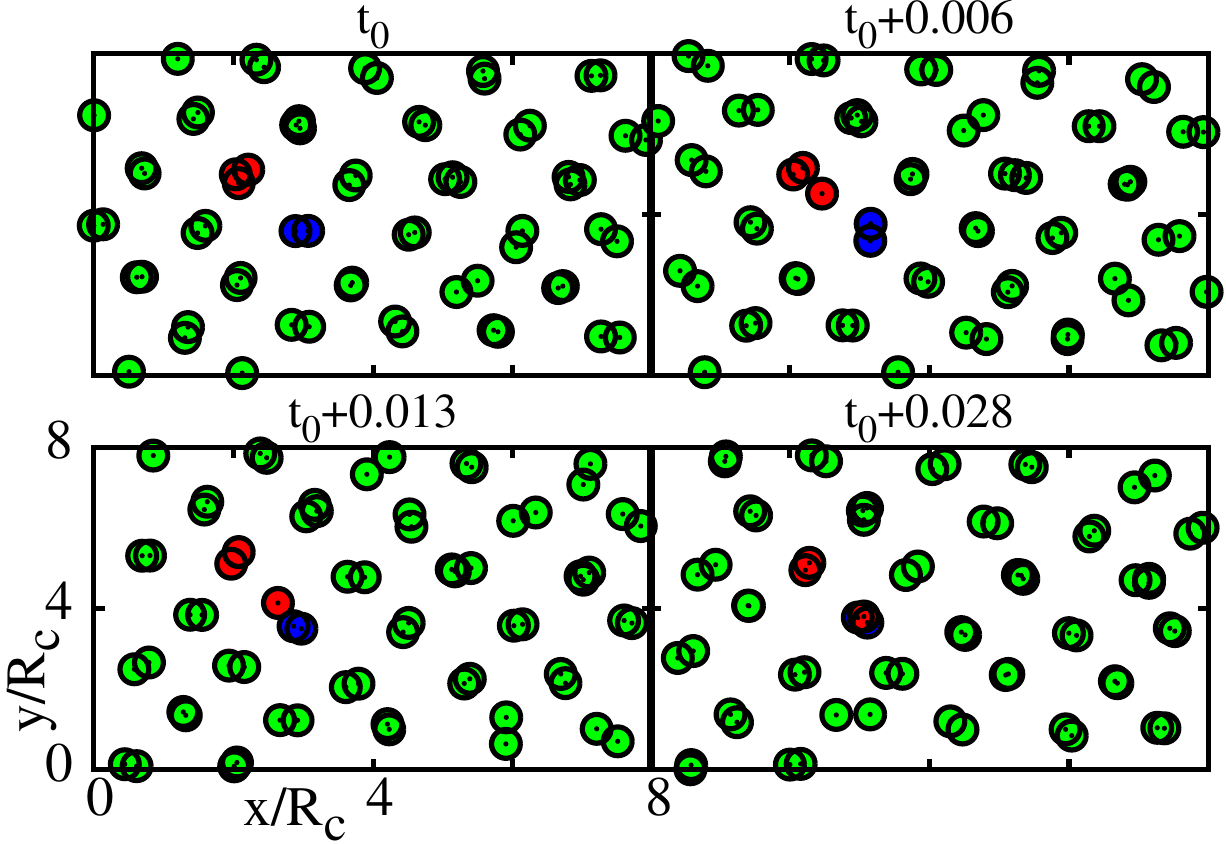} 
\caption{(color online) Snapshots of the sequential evolution of a 
portion of the system, starting at $t=t_0$. Particles
belonging to
sites involved in 
a hopping process have been highlighted in  red and blue.}
\label{fig:fig4}
\end{figure}

\subsection{Dynamical properties}
{\it The dynamics} of the equilibrium phases of Fig.~1 is instead initially characterised by computing 
the mean square displacement 
\begin{equation}
MSD(t)= \langle\Delta r^2(t)\rangle=\frac{1}{N}\langle \sum_j|\mathbf{r}_j(0)-\mathbf{r}_j(t)|^2 \rangle,
\label{eq:msd}
\end{equation}
as well as
the time-dependent
self-intermediate
scatter function 
\begin{equation}
F_s(k^*,t) = \frac{1}{N}\langle\sum_j e^{i k^* [{\bf r}_j(0)-{\bf r}_j(t)]}\rangle.
\label{eq:f_k}
\end{equation} 
Here,  $k^*=|{\bf k}^*|$ refers to the
characteristic wave vector of the main peak in $S(\mathbf k)$ \cite{LZ2013}. These quantities provide complementary
information on particle mobility and time-correlations of particle positions within the crystal, respectively \cite{liquids}:
For example, in the liquid phase, $MSD(t)$ follows the 
linear diffusion law $MSD(t) \propto t$ \cite{Ngai2011} typical of Brownian motion, while $F_s(k^*,t)$ decays exponentially. 

For model Eq.~(\ref{eq:ham}) the situation is however strikingly different. 
For $T\lesssim T_{\rm M}$ the time evolution of both observables, shown in Fig.~\ref{fig:fig3}(a) and (b), interpolates between the solid and liquid regimes: an extended plateau is followed by linear
diffusion for $MSD(t)$ and exponential decay for $F_s(k^*,t)$, respectively, with the size of the plateaux increasing with decreasing $T$. We find that an analysis of the $T$-dependent relaxation time $\tau_\alpha$ for which $F_s(k^*,\tau_\alpha)=1/e$ 
reveals an Arrhenius-type exponential dependence on $1/T$ without detectable saturation [see, e.g., Fig.~\ref{fig:fig3}(c)]. This indicates that the liquid-like diffusion described here is thermally activated, and only vanishes at $T=0$. 

These results are interesting as the phenomenology reported here, that is the existence of plateaux in the time evolution of 
$MSD(t)$ and $F_s(k^*,t)$, is typical  of e.g.,  glass-forming liquids. In particular, the Arrhenius-type dynamics would usually correspond to a peculiar ``strong-glass" behaviour. This immediately calls for an explanation of the corresponding microscopic dynamics responsible for this macroscopic behavior. 
\\ 

By inspection of particle configurations, we determine that the microscopic diffusion mechanism here corresponds to hopping of particles between different clusters, leaving the underlying crystal structure essentially unaltered. Examples of this dynamics are given in the snapshots of Fig.~\ref{fig:fig4}. These behaviors indicate the existence of {\it liquid-like particle diffusion within the crystalline phase at equilibrium}.  Specifically, for the case of our interest, namely $R_c^2 \rho >1$ , this hopping mechanism occurs in the presence of bond orientational order of the clusters for any temperature within the interval $0\lesssim T \lesssim T_{\rm M}$.\\

\begin{figure}[t]
\centerline{\includegraphics[width=1.0\columnwidth]{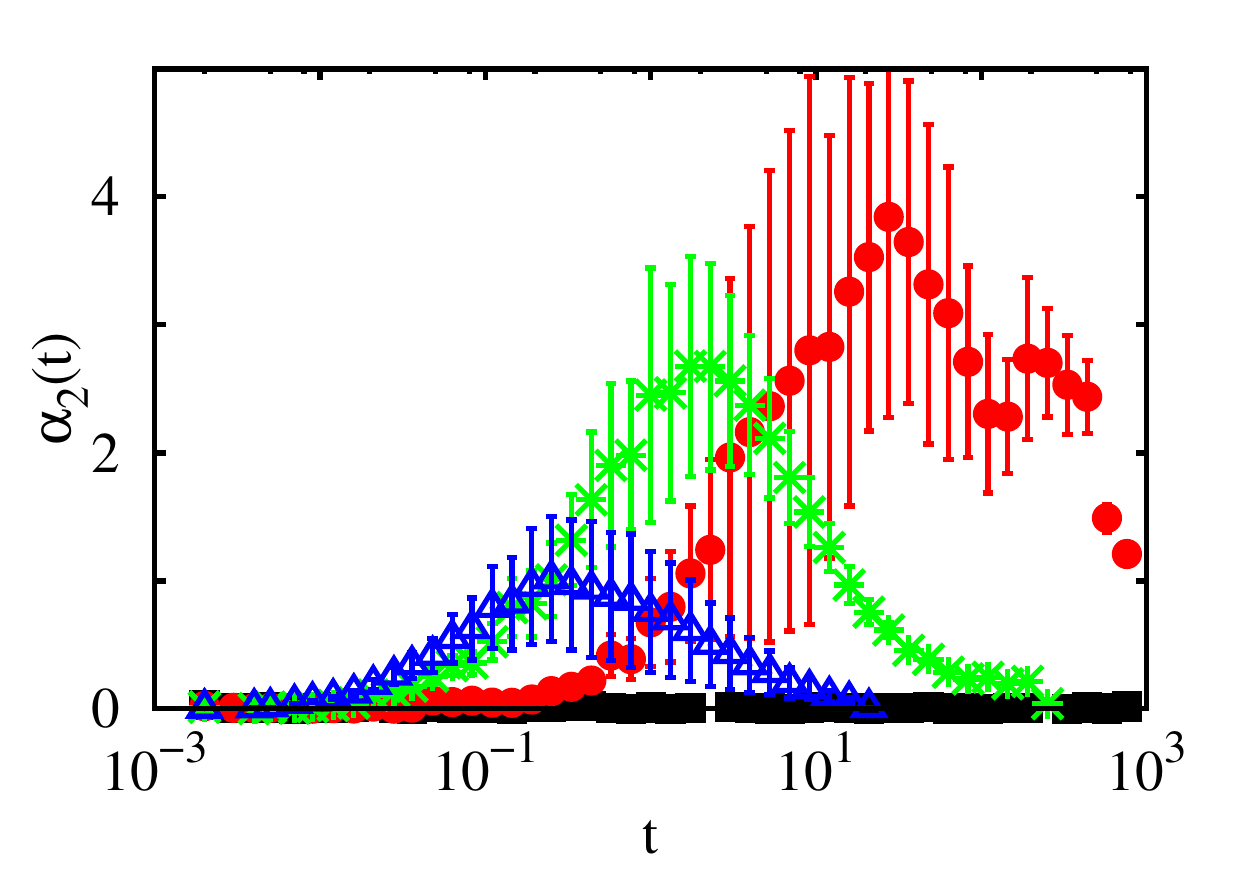}}
\caption{(color online) Non-gaussian parameter $\alpha_2(t)$ $vs.$ $t$ for
$T/(V_0/R_c^6)=0.03$ (black squares), $0.05$ (red circles), $0.07$ (green stars), $0.08$~(blue triangles).
Data refer to scaled density $R_c^2\rho=1.1$. }
\label{fig:fig5}
\end{figure}

We further characterize the dynamical properties of our system by estimating the following non-gaussian parameter \cite{ZAMPONI,Fyn2004}
\begin{equation}
\alpha_2(t)=\left[\frac{\langle\Delta r^4(t)\rangle}{2\langle\Delta r^2(t)\rangle^2}\right]-1.
\label{eq:alpha}
\end{equation}
The latter measures
deviations from gaussian fluctuations in the distributions of displacements, and thus is in general $\alpha_2(t)\simeq0$ for all
$t$ in regular liquids and non-cluster crystals at equilibrium.
Here, however, for $T<T_{\rm M}$ and intermediate times we obtain $\alpha_2(t) \neq 0$ since the particles can be differentiated in {\it fast} and {\it slow} due to deconfined inter-cluster hopping and confined intra-cluster motion. This is demonstrated for our model in Fig.~\ref{fig:fig5}, where, as before, all simulations have been performed at equilibrium. The presence of a peak in the non-gaussian parameter mimics, in our model, the so-called {\it dynamic
heterogeneity} usually associated with glassy dynamics where different time scales emerge. \\

We note that some of the general phenomena described above, such as, e.g., the deviation from gaussian distribution in the fluctuations of displacements, are characteristic of all cluster forming models, and have been recently observed numerically in 3D situations of relevance to soft-matter~\cite{csk11}. In addition to analysing the macroscopic order parameters and the specific microscopic realisation of these effects in a 2D model of direct experimental interest for low-temperature atomic physics, here we explain how (and in what sense) these cluster forming crystals correspond to the classical counterpart of the recently discovered cluster supersolids in the quantum regime. In fact, although the existence of a high-temperature crystalline phase related to the  supersolid phase was discussed previously~\cite{Cinti2010,Cinti2014}, its explanation in terms of dynamical properties is one of the main results of this work.\\

In the following, we start  bridging the gap between these classical and quantum regimes, by investigating quantum effects in the semiclassical and fully quantum cases, at equilibrium. 
Specifically, we show that while zero-point motion results in an enhancement of liquid-like properties, bosonic quantum statistics favors crystalline ones. The analysis of these separate, competing roles is crucial to understanding how the supersolid phase emerges from its classical analog.

\section{Quantum Effects}\label{Sec:Quantum}
{\it Quantum effects}  ($\hbar \neq 0$, $T>0$) on the phase diagram of Fig.~1 are investigated numerically in the semiclassical approximation using Path Integral Langevin Dynamics (PIMD) \cite{MARKLAND2011} and fully  quantum mechanically using exact quantum Path Integral Monte Carlo methods (PIMC) \cite{Boninsegni2006}. These provide complementary information: PIMD neglects particle exchange and the classical limit is the real-time dynamics of the particles, while PIMC treats the bosonic statistics exactly in imaginary time.\\

\subsection{Semiclassical dynamics}
\begin{figure}[b]
\centerline{\includegraphics[width=1.0\columnwidth]{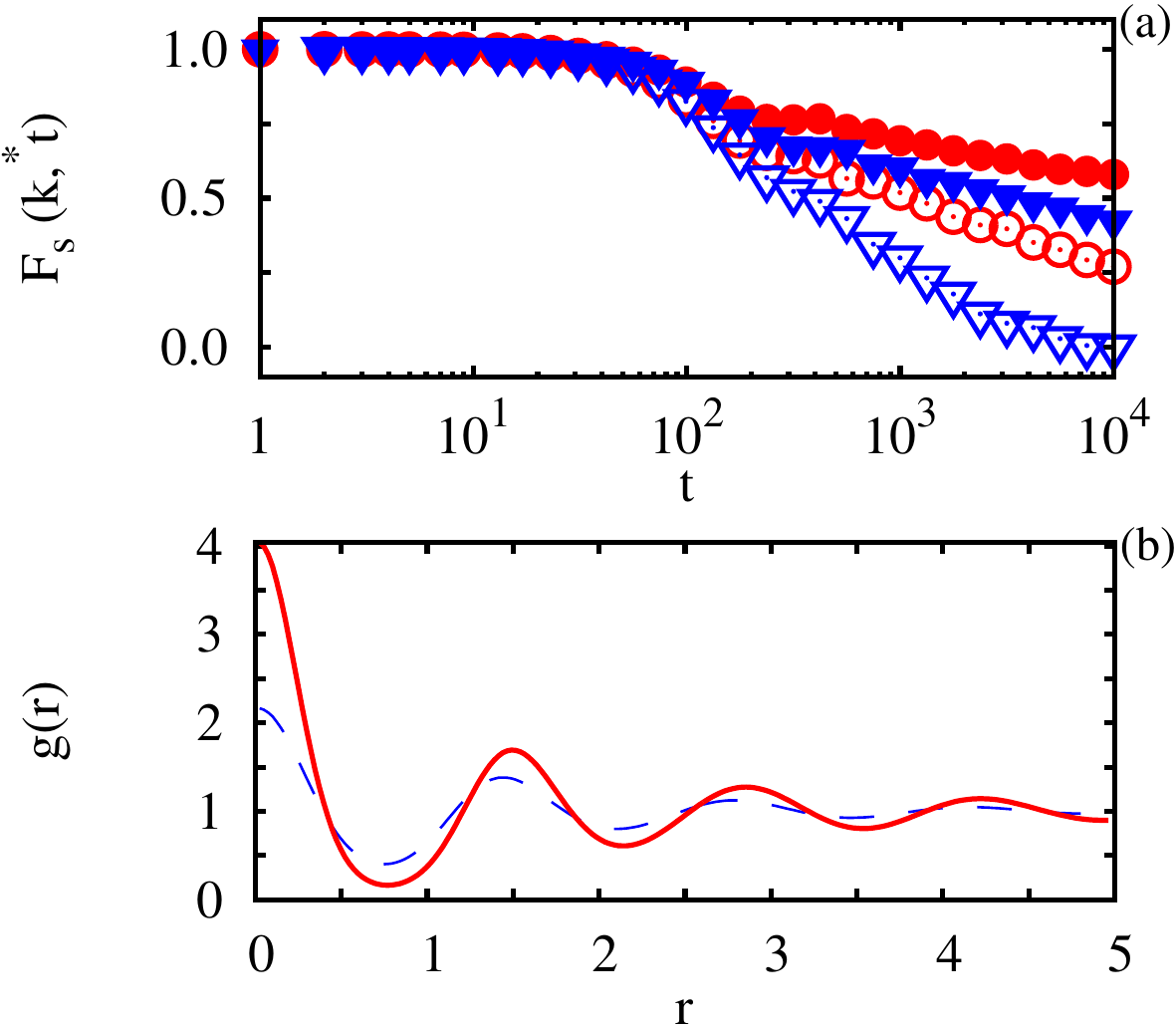}}
\caption{(color online) (a): $F_s(k^*,t)$ computed by Path Integral Langevin Dynamics. Data  are for $R_c^2\rho=0.8119$,
$mV_0/(\hbar^2R_c^4)=80$ (circles) and 60 (triangles). $T$ values are 1.6 and 2.6 (full and empty symbols, respectively), in
units of $\hbar^2/(m R_c^2)$. (b): $g(r)$ computed via Path Integral Monte Carlo simulations with and without bosonic
quantum exchanges (solid and dashed line, respectively). Here $T/ [\hbar^2/(m R_c^2)]=10$,  $mV_0/(\hbar^2R_c^4)=35$, and
$R_c^2\rho=2.029$.
}
\label{fig:fig6}
\end{figure}

Figure~\ref{fig:fig6}(a) shows results for $F_s(k^*,t)$ defined above using PIMD. We choose here
scaled densities such as $\alpha \gtrsim \alpha_{cs-ss}$, so that the zero-temperature quantum phase is a cluster crystal even in the quantum regime. In full generality, Fig.~\ref{fig:fig6} shows that semiclassical quantum fluctuations (e.g., "zero-point motion") cooperate with thermal fluctuations to enhance local mobility. At low $T$, this effect tends to destabilize crystalline order. \\

Indeed, fluctuations can affect the equilibrium glassy-like dynamics discussed above leading to melting into a liquid phase. As an example, the figure shows that for $R_c^2\rho=0.8119$ and $T=2.6$ [in units of $\hbar^2/(m R_c^2)$]  the relaxation time sensibly decreases when the interaction strength is lowered from  $mV_0/(\hbar^2 R_c^4)=80$ to $mV_0/(\hbar^2 R_c^4)=60$. These semiclassical results should be compared to those of Ref.~\cite{MARKLAND2011} for a gas of polydispersed hard-sphere liquids, where for certain parameters quantum fluctuations may induce an increase of glassiness, and thus a re-entrant behavior, before melting. 

\subsection{Effects of bosonic statistics}
The effects of bosonic quantum statistics are shown in Fig.~6(b). There, we present results for the 
radial density-density correlation function defined as
\begin{equation}
g(r)=\frac{1}{N}\langle\sum_j [\delta n_j/(2\pi r\delta r)]\rangle,
\label{eq:gr}
\end{equation} 
where $\delta n_j$ is the number of particles at a distance between $r$ and $r+\delta r$ from particle $j$.
$g(r)$ is computed for Boltzmann particles (i.e., no particle
exchanges, dashed line) as well as for Bose particles (i.e., including bosonic statistics, solid line)~\cite{Note1}. As an
example, we choose $R_c^2\rho=2.029$, $mV_0/(\hbar^2R_c^4)=35$, and $T/ [\hbar^2/(m R_c^2)]=10$. 

The figure shows that the
density-density correlations in the semiclassical case are liquid-like. 
In particular, oscillations of $g(r)$ beyond the first correlation shell are strongly damped, and the value of $g(0)$ is
approximately 2. However, this latter value increases by a factor of about 2 when quantum statistics is taken into account
(solid line), i.e., in a fully quantum mechanical calculation. In addition,  $g(r)$ displays more pronounced oscillations at
finite $r$, signaling the enhancement of solid-like  behavior. While less efficient for $R_c^2 \rho < 1$, we find that this
enhancement of solid-like properties is a general feature at sufficiently high density ($R_c^2\rho \gtrsim 1$) 
\cite{Cinti2014a}. This is in contrast to the physics of non-cluster crystals, such as purely dipolar bosons
\cite{bdl07,Astra07} or He$^4$ 
\cite{Boninsegni2012}, where, as shown in 
Ref.~\cite{QUANTUMJAMMING}, bosonic statistics always enhances superfluid properties.\\

In summary, from the analysis above we conclude that, surprisingly, the effects of quantum fluctuations and statistics can be competing: zero-point motion tends to destabilize the cluster crystal in favor of liquid-like phases,
while bosonic quantum statistics can have the opposite effect of enhancing crystalline behavior. Since PIMC simulations (as well as essentially any other methods) do not allow for the investigation of the real-time evolution of the interacting many-body problem in the presence of quantum exchanges, it remains an open question to determine the precise  dynamics of these systems in the quantum regime.

\section*{CONCLUSIONS AND OUTLOOK}\label{Sec:Outlook}
 We have demontrated that a two-dimensional model of monodisperse cluster forming particles can realize a classical equilibrium phase that simultaneously breaks both translational symmetry and dynamic homogeneity. While the latter phenomenon is usually associated with out-of-equilibrium glassy physics, here we find it at equilibrium. This results in the realization of a classical self-assembled cluster crystal with coexisting liquid-like properties. This corresponds to a classical analog of the quantum mechanical supersolid phase of matter.\\
 
The coexistence of a cluster crystalline structure and of particle diffusion has been here explained in terms of a thermally activated hopping mechanism, where particles delocalize without altering the underlying cluster crystalline matrix.   In addition, we have determined the competing effects that quantum mechanical fluctuations and statistics produce on a classical cluster crystal. As ultra-soft interactions are now observable in experiments with Rydberg atoms, this work may open up the exciting possibility of observing soft-matter phenomena in the classical and quantum regimes in atomic physics. Furthermore, interesting theoretical questions are still open, including whether out-of-equilibrium quench dynamics can suppress the structural order presented in this work, and what would be the resulting classical and quantum phases. 

\section*{ACKNOWLEDGMENTS}
We thank T. Pohl for discussions. W. L. acknowledges support by the Austrian Science Fund through Grant No. P 25454-N27 and by the Institut f\"ur Quantenoptics und Quanteninformation. G. P. acknowledges  support by the European Commission via ERC-St grant ColdSIM (No. 307688), EOARD, and UdS via Labex NIE and IdEX,  ITN COHERENCE, RYSQ, computing time at the HPC-UdS, ANR through "BLUESHIELD".

\section*{APPENDIX}\label{APP}
\section{Clustering technique}
\begin{figure}[ht]
\includegraphics[width= 1.0\columnwidth, angle=0]{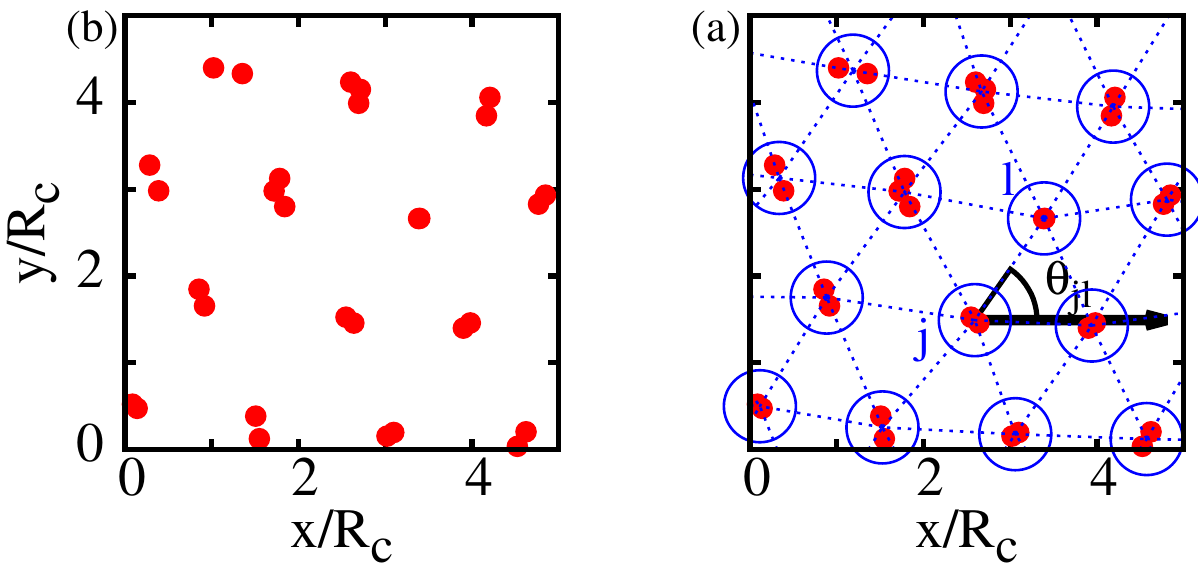} 	
\caption{(color online) Snapshot of a portion of the system in a cluster crystal configuration. (a): individual particles. (b): schematic outcome after the hierarchical clustering technique; blue circles are centered in cluster
centroids, dashed blue 
segments
connect nearest 
neighboring clusters, 
and $\theta_{jl}$ is the angle between the $x$ axis and the line
joining cluster $j$ with its neighbor $l$.}
\label{fig:fig7}
\end{figure}

In order to calculate the hexatic (short-range) order of the cluster crystal the first step is to 
distinguish
between different clusters. 
Here, we used a hierarchical clustering technique \cite{TANBOOK} 
that associates each particle to a single cluster in an unambiguous
way (Fig.~\ref{fig:fig7} and text below).  

For a given configuration of the system, the algorithm starts with $N_c=N$ one-particle clusters, corresponding to the $N$ single particles and their positions. Then, an iterative step consists in finding the minimum distance between all pair of
clusters, 
in order to merge
the two nearest clusters into a single one, and in relabeling the corresponding particles. The position of the new
cluster (formed by the union of the previous two) is defined as the centroid of all the associated particles. 
The procedure ends when the minimum distance between pairs of clusters is greater than a fixed number $d_c$. The value of $d_c$ has been set to $d_c=0.7 R_c$ in our calculations, roughly corresponding to half the value of the first
peak in the density-density correlation function. This peak remains at the same position for all densities.

\end{document}